\documentclass[final,5p,times,twocolumn,authoryear]{elsarticle}

\usepackage{amssymb}

\usepackage{amsmath}

\usepackage{color}

\journal{Quaternary International}

\renewcommand{\v}[1]{\mathbf{#1}}	

\begin{document}

\begin{frontmatter}



\title{On the roles of hunting and habitat size on the extinction of megafauna}


\author[fiestin,ib]{Guillermo Abramson}
\ead{abramson@cab.cnea.gov.ar}
\author[fiestin]{Mar\'{i}a F. Laguna}
\ead{lagunaf@cab.cnea.gov.ar}
\author[fiestin,ib]{Marcelo N. Kuperman}
\ead{kuperman@cab.cnea.gov.ar}
\author[fb,rio]{Adri\'{a}n Monjeau}
\ead{amonjeau@fundacionbariloche.org.ar}
\author[iidivca]{Jos\'{e} L. Lanata}
\ead{jllanata@conicet.gov.ar}

\address[fiestin]{Centro At\'{o}mico Bariloche and CONICET, R8402AGP Bariloche, Argentina}
\address[ib]{Instituto Balseiro, R8402AGP Bariloche, Argentina}
\address[fb]{Fundaci\'{o}n Bariloche and CONICET, R8402AGP Bariloche, Argentina}
\address[iidivca]{Instituto de Investigaciones en Diversidad Cultural y Procesos de Cambio, CONICET-UNRN, R8400AHL Bariloche, Argentina}

\begin{abstract}
We study a mechanistic mathematical model of extinction and coexistence in a generic hunter-prey ecosystem. The model represents typical scenarios of human invasion and environmental change, characteristic of the late Pleistocene, concomitant with the extinction of fauna in many regions of the world. As a first approach we focus on a small trophic web of three species, including two herbivores in asymmetric competition, in order to characterize the generic behaviors. Specifically, we use a stochastic dynamical system, allowing the study of the role of fluctuations and spatial correlations. We show that the presence of hunters drives the superior herbivore to extinction even in habitats that would allow coexistence, and even when the pressure of hunting is lower than on the inferior one. The role of system size and fluctuating populations is addressed, showing an ecological meltdown in small systems in the presence of humans. The time to extinction as a function of the system size, as calculated with the model, shows a good agreement with paleontological data. Other findings show the intricate play of the anthropic and environmental factors that may have caused the extinction of megafauna.   
\end{abstract}


\end{frontmatter}


\section{Introduction}
\label{}

The extinction of megafauna in many habitats of the world was coincidental with the dispersal of modern humans during the Pleistocene. Simultaneously, profound environmental changes took place as a consequence of the advance and retreat of repeated glaciations. A particularly interesting scenario is the one that took place in the Americas. Over a relatively short period of time during the Last Maximal Glaciation (LMG) this land was invaded by humans \citep{lanata2008,bodner2012}, who produced a lasting impact in many ecosystems. Actually, the occupation of the Americas and the widespread extinction of fauna was just one more act in a very long lasting play that accompanied the expansion of modern humans since their departure from Africa, and which ended in the 19th century with the colonization of the few remaining islands still free of our species. 

In our sister paper in this issue, \citet{monjeau2015} analyzed the ``controversy space'' of the debate regarding the causes of megafaunal extinctions in the Quaternary. As shown in the literature reviewed, common grounds and focus of discussions were changing throughout time, and notoriously, there is still no consensus about the ultimate cause of extinctions. Focused since 1966 in a passionate debate between climate versus human overkill \citep{koch2006,grayson2001}, the controversy space is suffering a conceptual blockage. This is because there is evidence supporting both hypotheses \citep{barnosky2004,lyons2004}. As \citet{ripple2010} wrote, it is more interesting to investigate what the role of humans might have been rather than debate solely the merits of overkill versus climate hypotheses. The extreme polarization between climate versus overkill expelled from the mainstream other causal hypotheses that have played important roles in the explanation of extinctions when the man was not involved, such  us predation, competitive exclusion and the role of area in the causal models, as if these biological effects had disappeared from the wild in the presence of man. Only recently, some biological arguments have reappeared in the debate. In addition, some recent attempts toward a multicausal synthesis have appeared in the literature (See \citet{monjeau2015} in this issue, Table 1, and references cited therein).
 
One way to solve the conceptual blockage in the controversial space may be, as we propose in \citet{monjeau2015}: to build a mechanistic mathematical model to evaluate the influence of each variable in different scenarios, so that supporters of either theory have that tool to resolve their disputes.

The mathematical modeling of the system, as it is generally recognized, might contribute in several ways to the understanding of a multiplicity of aspects. Considerable work has been devoted to it, with emphasis in different mechanisms with the purpose of dealing with a variety of details. The hunting hypothesis has been studied both analytically and numerically by \citet{flores2014} and by \citet{alroy2001}, among others. A review of several models with different points of view has been published by \citet{barnosky2004}.

The formulation of a complete mathematical model of the whole ecosystem of a continent  during its invasion by the first human populations is a daunting task. It would require not only modeling the densities of all the species, extended in space in their corresponding habitats, but also a detailed assessment of their interactions. Even if one would be able to do it to some extent, as pointed out by~\citet{tilman1987}, a phenomenological model taking account of the pairwise interactions between species would be insufficient in the case of a complex assemblage. The situation could be even more complicated by the existence of many-species interactions. On the one hand, the analysis of such a model would obscure the emergence of indirect effects from the interaction of its subsystems. On the other hand, it would ignore the role of spatial correlations that would be expected to develop in a stochastic and spatially extended landscape \citep{laguna2015}. 

We propose, for the reasons above mentioned, a mechanistic approach to the mathematical modeling of the system. This involves the specification of a set of axioms that determine the temporal evolution of the system from one state to the following. These axioms must embody information about the life history of the species involved (in the sense of~\citet{tilman1987}). Let us say that the state of the system is specified by a multidimensional vector $\v{u}$, with each component $u^\alpha$ corresponding to a relevant variable in the ecosystem: all the species, but also all the resources and physical variables  (water, shelter, space, etc.). Furthermore, let us suppose that the system is extended in a discrete space and denote  $\v{u}_i$ the value of the state vector at each site $i$. Formally, such a program should be cast in a model of the form:
\begin{equation}
\v{u}_i(t+\delta t) = \v{F}(\v{u}_{n(i)},t),
\label{topmodel}
\end{equation}
where $n(i)$ denotes a neighborhood of the site $i$ (including the site $i$ itself). The neighborhood of $i$ is relevant in the movement of species across the landscape through migration, colonization, invasion, etc. The key to getting some results from model (\ref{topmodel}) relies on two tasks: a good choice of the components of $\v{u}$ and a careful definition of the function $\v{F}$ that governs the temporal evolution. 

As said, an exhaustive specification of the system is a practical impossibility. Moreover, its analysis would barely give any insights into the phenomena of interest: the conditions for coexistence or extinction, for example. A drastic simplification of model (\ref{topmodel}) is imperative, with the purpose of turning it into a useful tool and a formal framework, with which such an analysis might be attempted. A reasonable starting point is to separate some of the components of $\v{u}$ and to consider them on their own. In \citet{monjeau2015}, Fig 1, we separated the controversial space in three types of causal explanations: biological, environmental and anthropic. Here we mirror this classification to build the model parameters. 

\begin{figure}[t]
\centering 
\includegraphics[width=\columnwidth, clip=true]{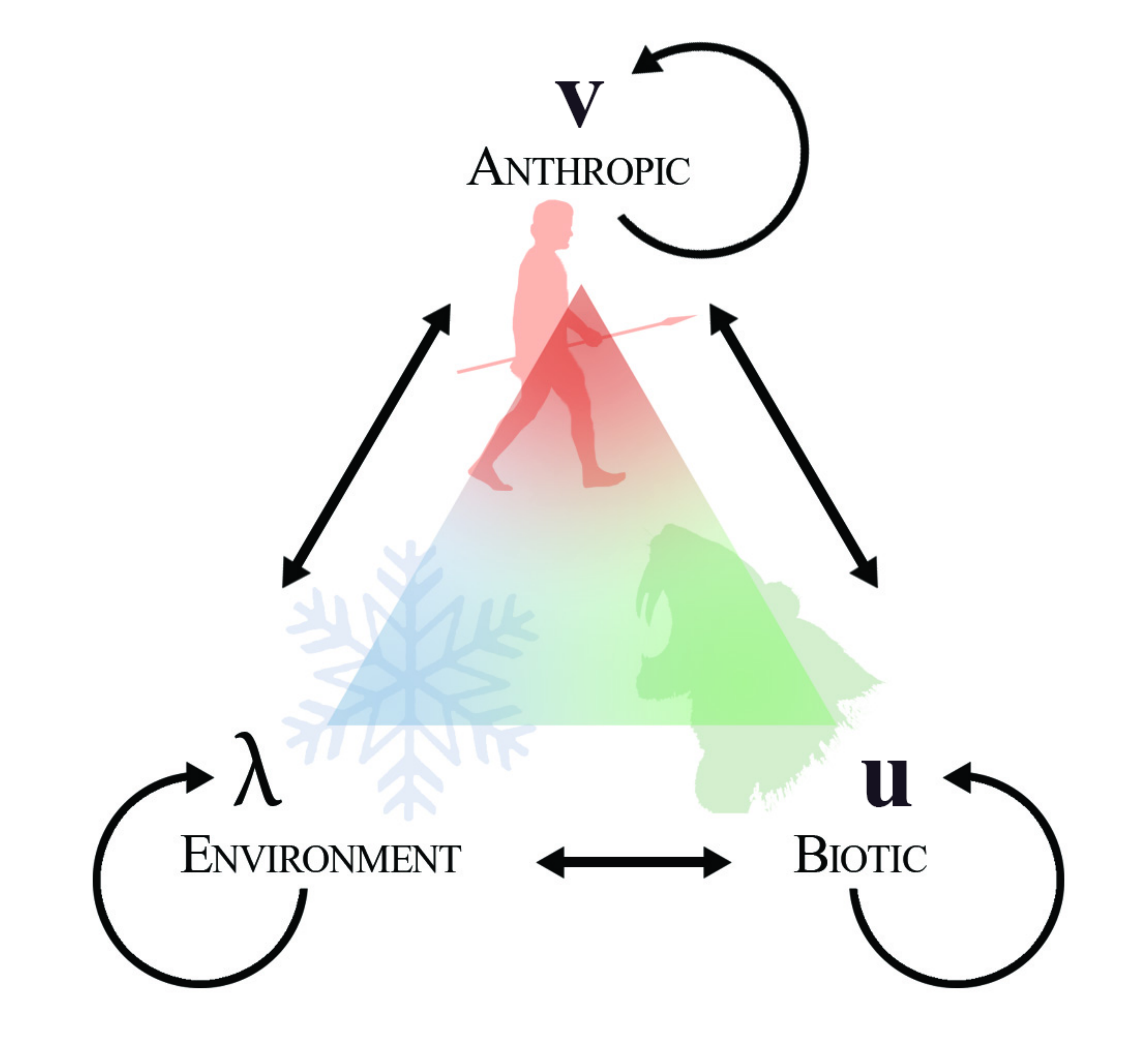}
\caption{Conceptual model of three variables, representing the biological, anthropic and physical/environmental factors described by model (\ref{transmodel1}-\ref{transmodel3}). The arrows represent the functions that govern the dynamics of the system.}
\label{uvlambda}
\end{figure}

Firstly, let us distinguish the biological \emph{variables} from the \emph{parameters} of the model, and represent the latter collectively by the vector $\boldsymbol{\lambda}$. These can be the physical elements of the system, but also any environmental parameters. An example is the size of the available habitat, which will play a role below. Secondly, let us also separate from $\v{u}$ the set of anthropic variables (let us call them $\v{v}$), leaving in $\v{u}$ just the biotic non-human ones: 
\begin{align}
\v{u}_i(t+\delta t) &= \v{F}(\v{u}_{n(i)},\v{v}_{n(i)},\boldsymbol{\lambda}_{n(i)},t), \label{midmodel1}\\
\v{v}_i(t+\delta t) &= \v{G}(\v{u}_{n(i)},\v{v}_{n(i)},\boldsymbol{\lambda}_{n(i)},t), \label{midmodel2}\\
\boldsymbol{\lambda}_i(t+\delta t) &= \boldsymbol{\Theta}(\v{u}_{n(i)},\v{v}_{n(i)},\boldsymbol{\lambda}_{n(i)},t),
\label{midmodel3}
\end{align}
where appropriate functions $\v{G}$ and $\boldsymbol{\Theta}$ give the evolution of the humans and the parameters respectively. A pictorial representation of such a model is given in Fig.~\ref{uvlambda}.

Further simplifications are necessary before model (\ref{midmodel1}-\ref{midmodel3}) can be used in practice. The first one is a separation of the dynamical evolution of the variables into a \emph{local} component (say, the demographics) and a \emph{transport} term. In addition, let us take all the parameters as independent of time. In our analysis below we will use them as control parameters to study different scenarios. Bear in mind that this assumption should be relaxed to allow the study of the full dynamics of the system. For example, the phenomenon of desertification by overgrazing would need a dynamical parameter describing the destroyed habitat coupled to the variable corresponding to the culprit herbivore. But this will be done elsewhere. The model becomes schematically:
\begin{align}
\v{u}_i(t+\delta t) &= \v{F}_0(\v{u}_i,\v{v}_i,\boldsymbol{\lambda}_i)
+\v{F}_1(\v{u}_{n(i)},\v{v}_{n(i)},\boldsymbol{\lambda}_{n(i)}) 
\label{transmodel1} \\
\v{v}_i(t+\delta t) &= \v{G}_0(\v{u}_i,\v{v}_i,\boldsymbol{\lambda}_i)
+\v{G}_1(\v{u}_{n(i)},\v{v}_{n(i)},\boldsymbol{\lambda}_{v(i)}),
\label{transmodel2} \\
\boldsymbol{\lambda}_i(t+\delta t) &= \boldsymbol{\lambda}_i(t),
\label{transmodel3}
\end{align}
where $\v{F}_0$ and $\v{G}_0$ are now local, involving only the variables and the parameters at site $i$, while the functions $\v{F}_1$ and $\v{G}_1$ take care of the spatial coupling within the neighborhood $n(i)$. Any explicit time dependence has also been ignored, assuming that their eventual change occurs at a slower time scale than the one corresponding to the animal populations.

Model (\ref{transmodel1}-\ref{transmodel3}) is a good starting point to give a specification of the structure of the system in terms of its subsystems and their interactions. A good minimal model is a two-level trophic web, composed of two herbivores in competition with a common predator. The predator will represent the invading population of human hunter-gatherers, and we will analyze below the reaction of a preexisting equilibrium population of herbivores to their appearance in the system. In addition, the competing interaction between the herbivores will be supposed to be asymmetric, or hierarchical, as will be discussed. 

A final specification corresponds to the habitat and its spatial structure. We choose a framework which is particularly suited to capture the role of a structured habitat and hierarchical competition: metapopulation models, as introduced by~\citet{levins71} and successfully developed by~\citet{tilman94} and~\citet{bascompte96} among others. It is worth noting that some of these approaches are based on a continuous and analytic formalism (a mean-field, as it is sometimes called), describing the dynamics with the use of differential equations. Such is the case of \citet{flores2014}, with his predator-prey model of megafauna in a generalized Leslie formalism. We will follow a different approach, as will be argued below.  

We will analyze in this system a variety of scenarios of coexistence and extinction, considering different systems sizes and levels of habitat destruction. In particular, we show that the extinction of megafauna can occur even without overkill (a mechanism extensively analyzed by some researchers), i.e. when the smallest animals are the preferred game of the humans. In our model, extinction arises from the synergic roles of hunting and habitat deterioration affecting the biodiversity of the ecosystem, each one in a peculiar way. 

\section{Three-species metapopulation model}

Let us now proceed to the formal specification of the model, which requires the definition of the dynamical functions of Eqs.~(\ref{transmodel1}-\ref{transmodel3}). In this Section we will describe a stochastic implementation of the model, which we will analyze numerically. A mean field approximation, as a complementary description, is briefly discussed in Appendix A. As we said above, the system is composed of three characteristic species. The metapopulations occupies an $L\times L$ square arrangement of patches of suitable habitats. These patches constitute their only resource. They can be colonized or vacated according to rules to be specified. The patches can also be destroyed and unsuitable for colonization,  representing a spatial heterogeneity and fragmentation of the landscape, as in~\citet{bascompte96}. 

We suppose, as in~\citet{tilman94}, that the two competing herbivores are not equivalent. There is a superior competitor that can colonize any patch that is neither destroyed nor already occupied by themselves. In addition, the inferior herbivore competitor can only colonize patches that are neither destroyed, nor occupied by themselves, \emph{nor} occupied by the superior one. With the purpose of giving the model an additional flexibility, we will furthermore consider two variants of this relationship. There is a \emph{strong} interaction where the superior herbivore competitor can displace the inferior one from a colonized site (this is the one considered by~\citet{tilman94}), and a \emph{weak} one where this displacement does not occur. 

It is usual to assume that the hierarchical difference between the competitors corresponds to their body size, with the larger animal being the superior one. Indeed, such is the case in many real situations \citep{tilman94}, but there are other possibilities of equal interest. For example, slow breeding vs. fast breeding animals may also be the source of the asymmetry. The coexistence of asymmetric competitors require that the inferior one has \emph{some} advantage over the superior one \citep{tilman94,kuperman96}. Usually the superior competitors are worse colonizers, but they could be slower breeders instead. And, as it has been pointed out by \citet{johnson2002}, slow breeding animals were also hard hit in the extinction event. In the following we will refer to superiority or size in a loose way, while in a strict sense we just imply the hierarchical asymmetry of the competition. Regarding the two variants of competitive interaction, a strong one may represent the one existing between a large and a small species with a significant overlap of their diets. In such a case the inferior species may run or disappear from a patch where they compete with the larger species. On the contrary, if the diets barely overlap, both species might coexist in the same patch, yet the smaller species may be unable or reluctant to colonize a patch already occupied by the superior species. 

The dynamics of occupation and abandonment of patches is inspired by the different ecological processes that drive the metapopulation dynamics. Vacant patches can be colonized and occupied ones can be freed (in a manner that will be described below in more detail). Hunters (predators) can only colonize patches already occupied by game (prey) (as in~\citet{swihart01} or~\citet{srivastava2008}). Their effect will be taken into account as an increase in the probability of local extinction of a population of the herbivores in the presence of hunters~\citep{swihart01}. Observe that the benefit that the hunters obtain from their prey comes from the fact that they can only colonize space where game is available. This closes the feedback between hunters and prey in mutual dependence. 

The dynamical variables of the model are the occupation of patches, disregarding the actual population density, number of individuals or biomass present at each patch. We denote the fraction of patches occupied by herbivores of species $i$ as $x_i$ (with $i=1,\,2$ for the superior and inferior ones respectively). So, $\v{u}=(x_1,x_2)$. The vector describing the anthropic variable $\v{v}$ has a single component, $y$, the fraction occupied by the humans. We will take into consideration a single  environmental, parameter $\boldsymbol{\lambda}$: the arrangement of patches. The fraction of usable (non-destroyed) habitat, $H$, will be used as a control parameter of the different scenarios. A value $H=1$ represents a pristine habitat. Smaller values of $H$ may of course obey to different causes, such as global environmental changes. 

In the model the time advances discretely and at each time step the following stochastic processes can change the state of occupation of a patch, effectively defining $\v{F}_0$, $\v{F}_1$, $\v{G}_0$ and $\v{G}_1$ in an algorithmic way:

\textbf{Colonization.} An available patch can be colonized by the species $\alpha$ from a first-neighbor occupied patch, with probability of colonization $c_\alpha$ ($\alpha$ being $x_1$, $x_2$ and $y$). Note that the availability of patches must take into account the asymmetry of the hierarchical competition: Species 1 can colonize any undestroyed patch not occupied by themselves, while species 2 can only colonize undestroyed patches free of \emph{any} herbivore occupant. Humans can only colonize patches already occupied by herbivores.

\textbf{Extinction.} An occupied patch can be vacated by species $\alpha$ with probability of local extinction $e_\alpha$. 

\textbf{Predation.} A patch that is occupied by either prey and by humans has a probability of extinction of the prey, given by a corresponding probability $\mu_\alpha$ (note that $\mu_y=0$).

\textbf{Competitive displacement.} A patch occupied by both herbivores can be freed of the inferior one $x_2$ with probability $c_{x_1}$. Note that there is no additional parameter to characterize the hierarchy: the colonization probability of the higher competitor plays this role. This is the \emph{strong} version of the competitive interaction mentioned above; the \emph{weak} one does not include this process.

In the formalism described in the Introduction, the non-local functions $\v{F}_1$ and $\v{G}_1$ are described by the colonization processes of prey and hunter-gatherers, respectively. The function $\v{F}_0$, which stands for the local dynamics of the herbivores, comprises extinction, predation and eventually competitive displacement. Correspondingly, $\v{G}_0$ is taken into account by the extinction of humans. 

In Appendix A we analyze a mean field model of the dynamics just described. The interested reader may find there the corresponding mathematical formulation in terms of explicit functions of the densities $x_1$, $x_2$ and $y$. Such deterministic models do not present some of the phenomena that are relevant in our discussion (fluctuations, spatial correlations, etc.). For this reason we present here the results of numerical simulations of the stochastic dynamics. We perform computer simulations on a system enclosed by impenetrable barriers (see details in Appendix B). To perform a typical realization we define the parameters of the model and destroy a fraction $D$ of patches, which will not be available for colonization for the whole run. The available habitat $H$ is the fraction of patches that can be colonized by the three species, i.e., $H=1-D$. Then, we set an initial condition occupying at random 50\% of the  available patches for each herbivore species. Human occupation of a fraction of the patches already colonized by herbivores may or may not be set, as will be discussed in the Results section.  The system is then allowed to evolve synchronously according to the stochastic rules. At each time step, each patch is subject to the four events in the order given above (actually, the order is irrelevant due to the synchronicity of the update). In the following section we show results of individual runs as well as temporal and ensemble averages. In the first case, the value of the variables $x_1$, $x_2$ and $y$ are recorded as a function of time. In the second case, the system is run for a total of 60,000 time steps, recording the time of extinction of each species, it they occur. Multiple repetitions of these runs are used to compute ensemble averages of the probability and time of extinction. 

\begin{table}[t]
\centering
\begin{tabular}{|l|c|c|c|}
\cline{1-4}
Species        & Colonization           & Extinction             & Predation           \\ \cline{1-4}
palaeolama ($x_1$)& $c_{x_1}=0.02$ & $e_{x_1}=0.02$ &  $\mu_{x_1}=0.1$ \\ \cline{1-4}
mylodon ($x_1$)& $c_{x_1}=0.02$ & $e_{x_1}=0.02$ &  $\mu_{x_1}=0.1$ \\ \cline{1-4}
guanaco ($x_2$)  & $c_{x_2}=0.04$ & $e_{x_2}=0.008$ & $\mu_{x_2}=0.2$  \\ \cline{1-4}
human ($y$)      & $c_{y}= 0.02$   & $e_{y}= 0.01$   &  $\mu_{y}=0$ \\ \cline{1-4}
\end{tabular}
\caption{Value of the metapopulation parameters used in this work. The difference between palaeolama and mylodon lies in the competitive displacement, as described in Section 2.}
\label{tabla}
\end{table}

The values of the parameters were chosen following ecological considerations from the natural history of South America. We chose one of the herbivores representing the guanaco (\emph{Lama guanicoe}), which is (and was) a widespread camelid that has survived the human invasion of South America Southern Cone. For the second herbivore we have considered two alternatives separately: mylodon, a giant ground sloth (\emph{Mylodon sp.}) and palaeolama, a giant camelid (\emph{Palaeolama sp.}). Both are extinct genera of once widespread populations, which disappeared between 10,000 to 8000 years BP concurrently with the human invasion and the environmental changes that followed the end of the Pleistocene. Mylodon and palaeolama weighed 300 kg, about three times more  than guanaco, which body mass is about 90 kg \citep{farina2013}. Observations of extant herbivores show that larger herbivores are able to displace the smaller ones from water sources and shared territory if their diets overlap \citep{nabte2013}. In this spirit, we choose to model mylodons and palaeolamas at the higher place in the competitive hierarchy. There is, nevertheless, a fundamental difference in the interaction of guanacos with mylodons and palaeolamas: guanaco and palaeolama diets overlapped, whereas guanaco's and mylodon's did not \citep{heusser1994}. This situation is taken into account in the model by neglecting the competitive displacement term for the case of the mylodon (what we called weak competitive interaction above).  

On the other hand guanacos, being the inferior competitor, need to display some advantage in order to persist under these asymmetrical conditions. Inspired by the correlation existing between body size and the reproductive potential of extant herbivores \citep{johnson2002}, we assume that the reproductive potential of guanacos is twice the ones of palaeolama or mylodon.  In the model, a higher reproductive rate for the guanacos is taken in to account by setting a higher colonization rate and a lower extinction rate compared to the ones set for the superior herbivores. The values of the parameters used in the results shown below are presented in Table 1, and have been chosen so that extensive regions of coexistence are observed when varying the control parameters.

Even if we agree with \citet{johnson2002} in the general pattern, we think, however, that body size may have some influence in the rate of extinction when the availability of energy of the habitat is scarce, given that large species requires more area to maintain a viable population than the smaller ones \citep{rapoport1982}. This may be the case of extinction in islands, where even low rates of killing can have susbstantial impacts on large species because the island can only support a small number of individuals. Small numbers are subject to the laws of fluctuations in small systems and can drive the species to the extinction, as proposed by the broken zig-zag model for paleobiology (\citet{cione2003}; also see item 3.3. below). 

\begin{figure}[t]
\centering 
\includegraphics[width=\columnwidth, clip=true]{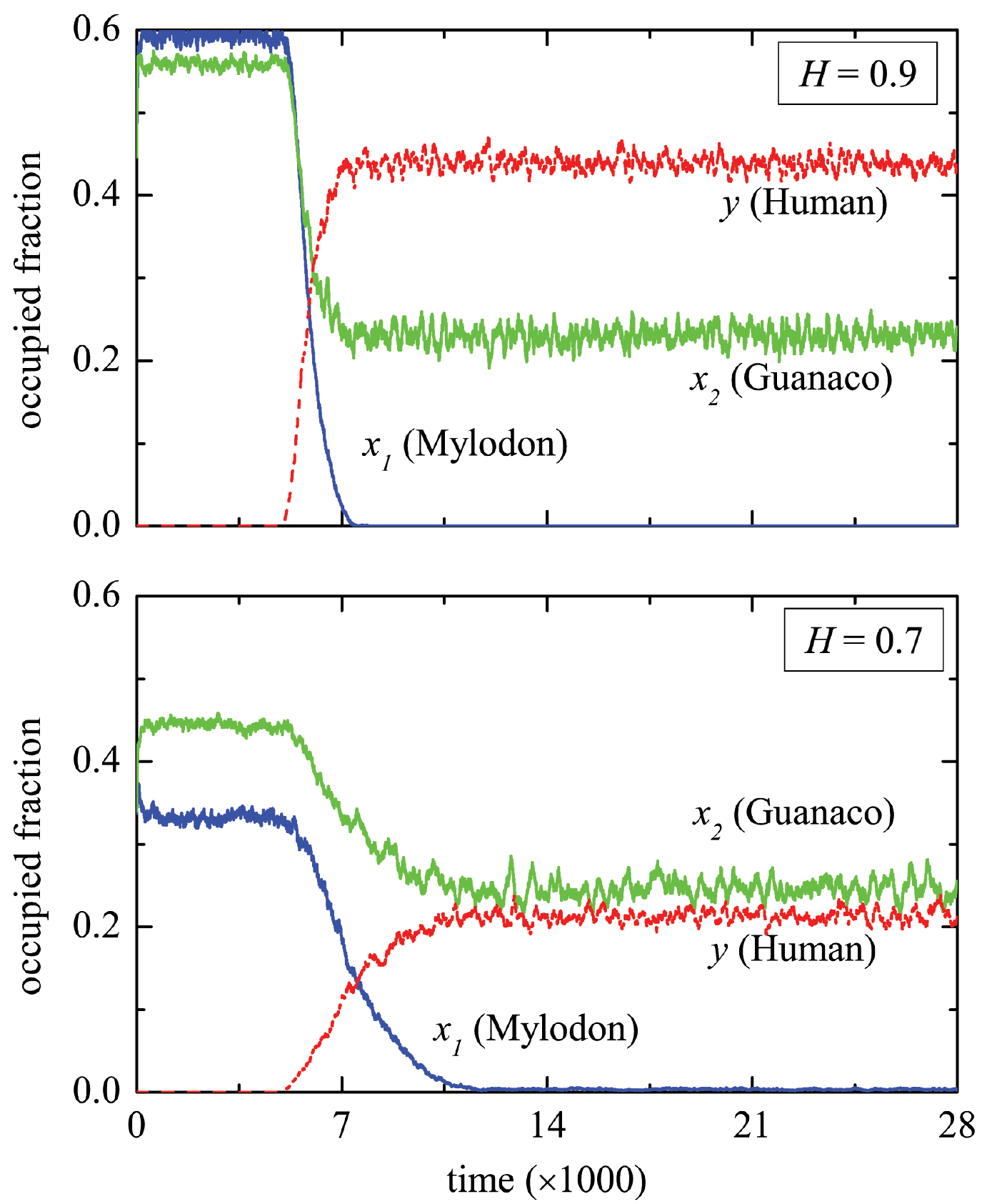}
\caption{Fraction of occupied patches of the three species (mylodon, guanaco and human) as a function of time for two values of available habitat, $H=0.9$ (upper panel) and $0.7$ (lower panel). During the first $5000$ time steps only the two herbivores $x_1$ and $x_2$ occupy the landscape. Afterwards humans are introduced in the system. The system size is $100 \times 100$, and 10 patches are occupied by humans at time $t=5000$. Other parameters as in Table 1.}
\label{milodon}
\end{figure}

\begin{figure}[t]
\centering 
\includegraphics[width=\columnwidth, clip=true]{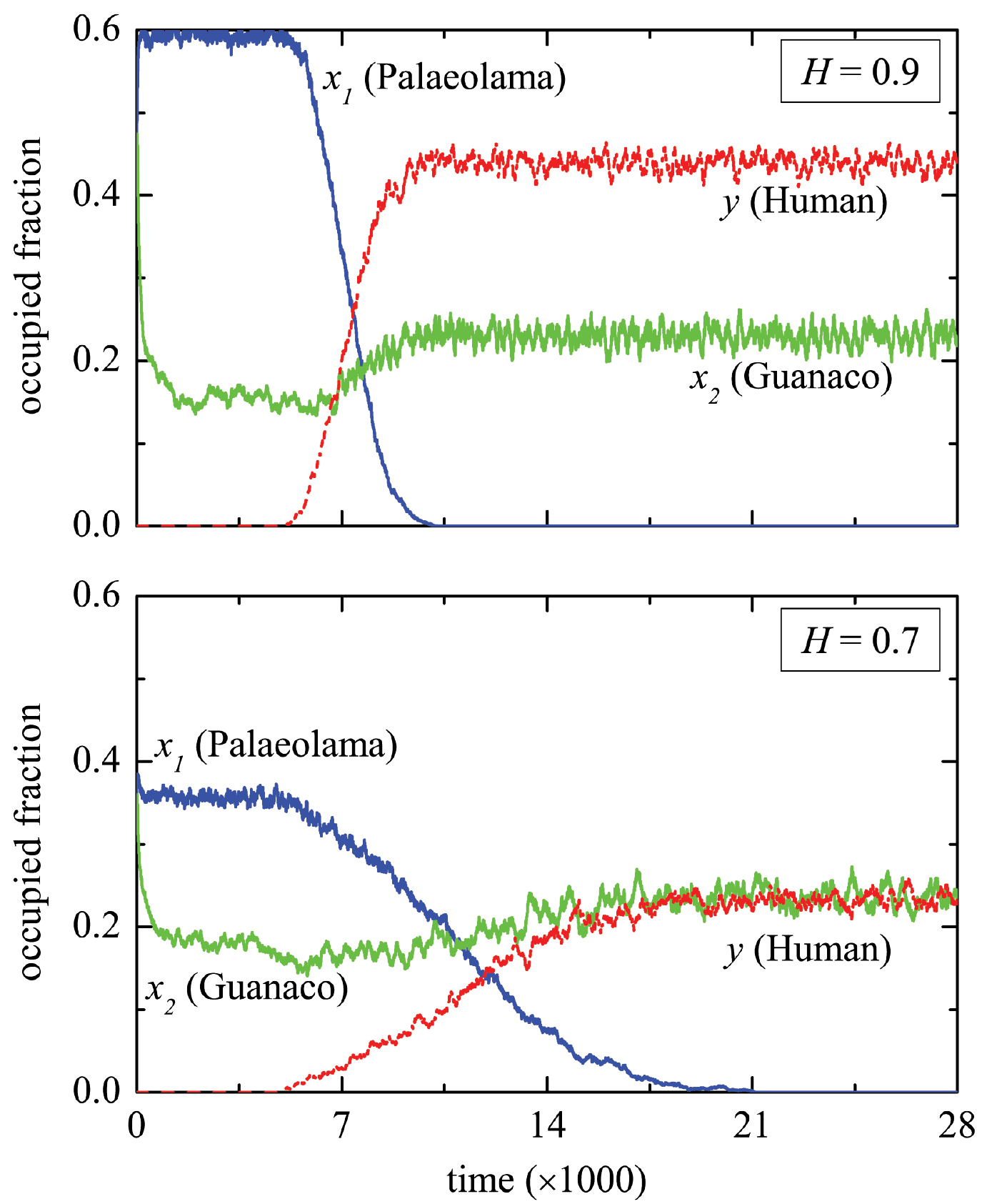}
\caption{Fraction of occupied patches of the three species (palaeolama, guanaco and human) as a function of time for two values of available habitat, $H=0.9$ (upper panel) and $0.7$ (lower panel). During the first $5000$ time steps only the two herbivores $x_1$ and $x_2$ occupy the landscape. After this time humans are introduced in the system. The system size is $100 \times 100$, and 10 patches are occupied by humans at time $t=5000$. Other parameters as in Table 1.}
\label{palaeolama}
\end{figure}

\section{Results}
\subsection{The human invasion}

In this Section we model a situation in which the two herbivores coexist without the presence of humans during an initial stage. After a given time (5000 time steps in the results shown below) humans are added to a small fraction of the patches occupied by the herbivores. 

In Fig.~\ref{milodon} we show results for two values of available habitat, $H=0.9$ and $0.7$ for the system mylodon-guanaco-human, while Fig.~\ref{palaeolama} corresponds to the system palaeolama-guanaco-human. In both figures, and for the two values of $H$, we observe that the herbivores coexist during the first stage in which humans are absent. The situation drastically changes when the humans are included in a very small number of patches. The humans (being predators) quickly colonize the patches occupied by the herbivores. After some transient time (which depends on $H$ and on the type of competitive interaction) the fraction of patches occupied by humans reaches a stationary value. We observe that the main consequence of the inclusion of humans is the extinction of $x_1$, the superior herbivore (palaeolama or mylodon). On the contrary, the inferior one ($x_2$, the guanaco) succeeds to survive, although it suffers a reduction of their occupied habitat.

In addition, the fraction of available habitat $H$ has two main effects: to determine the stationary fractions of each species and the rate of extinction of the superior herbivore $x_1$.

Moreover, strong and weak competition (i.e. with or without competitive displacement, palaeolama vs. mylodon cases) is only reflected in the ratio $x_1 / x_2$ during the stage where humans are absent ($t<5000$). The introduction of the hunter-gatherers in the system is much more determinant than the hierarchy between herbivores. Specifically, observe that the stationary state, which involves the same species (humans and guanacos), is the same in the two systems.

\begin{figure}[t]
\centering 
\includegraphics[width=\columnwidth, clip=true]{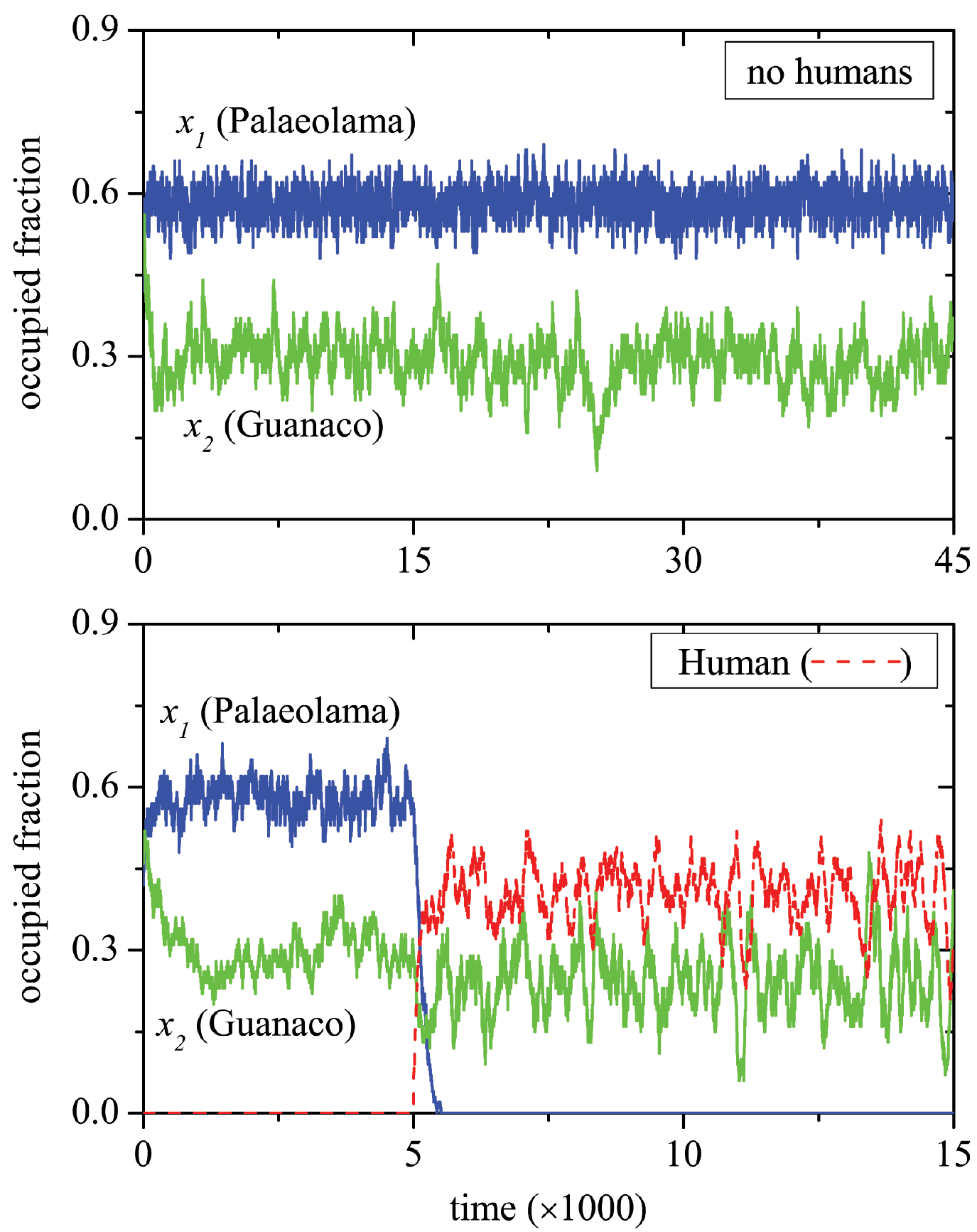}
\caption{Fraction of occupied patches of the three species (palaeolama, guanaco and human) as a function of time, for $H=0.9$ and a grid of $20 \times 20$ patches. In the upper panel the evolution of the two herbivores in absence of humans is shown. Lower panel: The case presented in the previous section. During the first $5000$ time steps only the two herbivores $x_1$ and $x_2$ are occupying the space. After this time, humans are introduced in the system. Six patches are occupied by humans at time $t=5000$. Other parameters as in Table 1.}
\label{L20}
\end{figure}

\begin{figure}[t]
\centering 
\includegraphics[width=\columnwidth, clip=true]{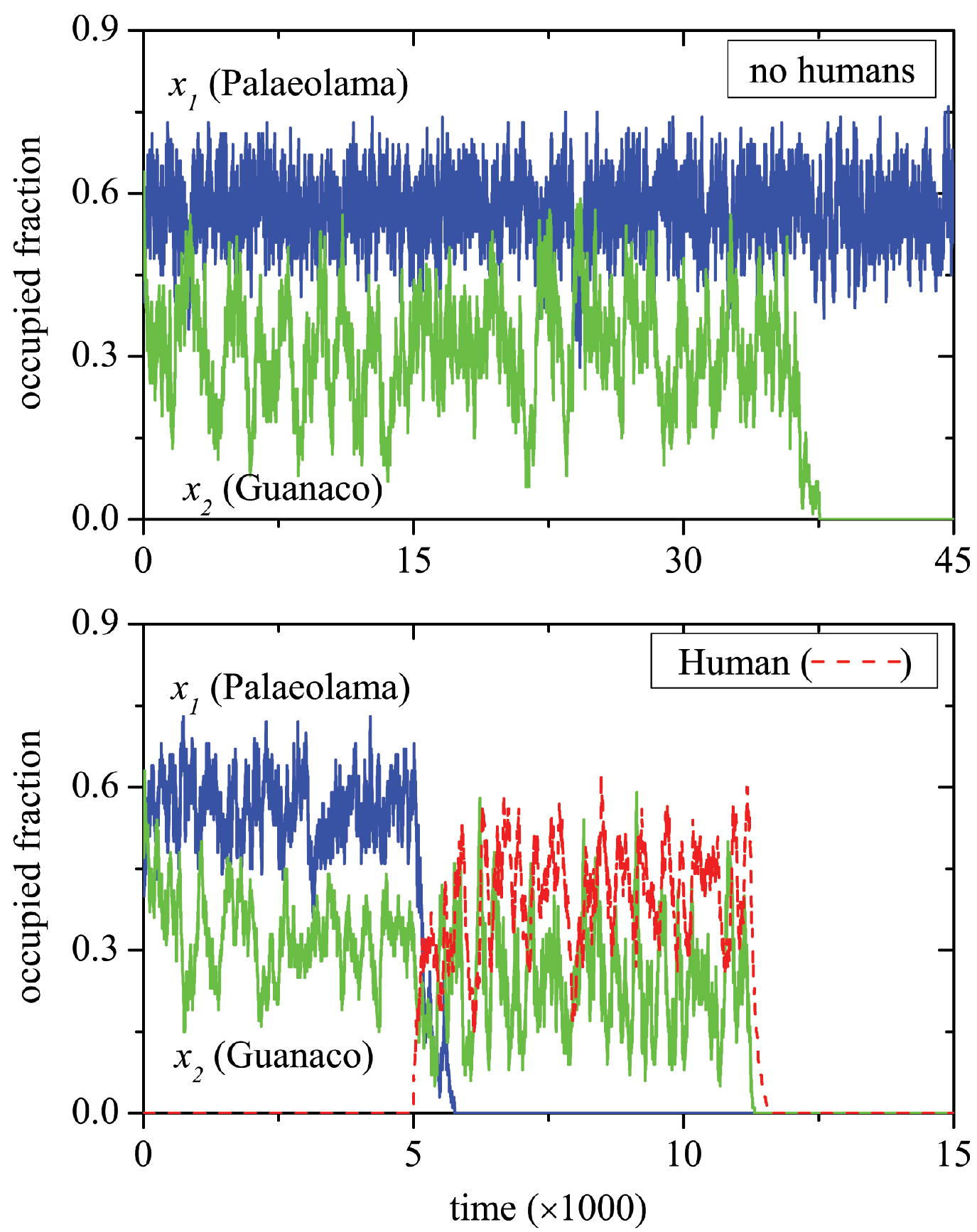}
\caption{Fraction of occupied patches of the three species (palaeolama, guanaco and human) as a function of time, for $H=0.9$ and a grid of $10 \times 10$ patches. In the upper panel the evolution of the two herbivores in absence of humans is shown. Lower panel: The case presented in the previous section. During the first $5000$ time steps only the two herbivores $x_1$ and $x_2$ are occupying the space. After this time, humans are introduced in the system. Three patches are occupied by humans at time $t=5000$. Other parameters as in Table 1.}
\label{L10}
\end{figure}

\subsection{Habitat size and species extinctions}

We analyze here the effect of the habitat size on species survival. For the sake of brevity, let us focus on the palaeolama-guanaco-human system, varying the habitat size (i.e., changing the size of the grid for a single value of the habitat availability $H$).  In Fig.~\ref{L20} we show results for the case $H=0.9$ (a quite pristine habitat) in a grid of $20 \times 20$, much smaller than the one presented in Fig.~\ref{palaeolama}. As a reference, in the upper panel we plot the temporal evolution of the system composed of the two herbivores only. Note that, in the absence of humans, both species coexist with a higher occupation of the palaeolama as it is the superior competitor. The lower panel shows the same case analyzed in the previous section, where humans are introduced at time $t=5000$. Their effect on the system is very similar to the one observed in the larger $100\times 100$ system. The most apparent difference is the enhanced amplitude of the fluctuations, a behavior expected for small systems.

A further reduction of the grid size produces a very different result. In Fig.~\ref{L10} we present a system of $10 \times 10$ patches. The case shown in the upper panel suggests that extinctions can take place in the absence of humans if the habitat size is small enough. Large fluctuations are responsible for this extinctions, that happen more frequently for the inferior competitor, $x_2$, as will become clearer in the following Section. When humans are introduced their effect on the herbivores is also different (see the lower panel of Fig.~\ref{L10}). On the one hand, the process of extinction of $x_2$ is accelerated. On the other hand, palaeolama is also driven to extinction and humans collapse a few time steps later. 

The behavior just described is not the only possible result of the simulations. Such small systems are prone to be governed by fluctuations, and as a consequence different realizations of the stochastic process produce different results. In the next Section we analyze how frequent is this scenario and how it changes with the value of $H$.   

\begin{figure}[t]
\centering 
\includegraphics[width=\columnwidth, clip=true]{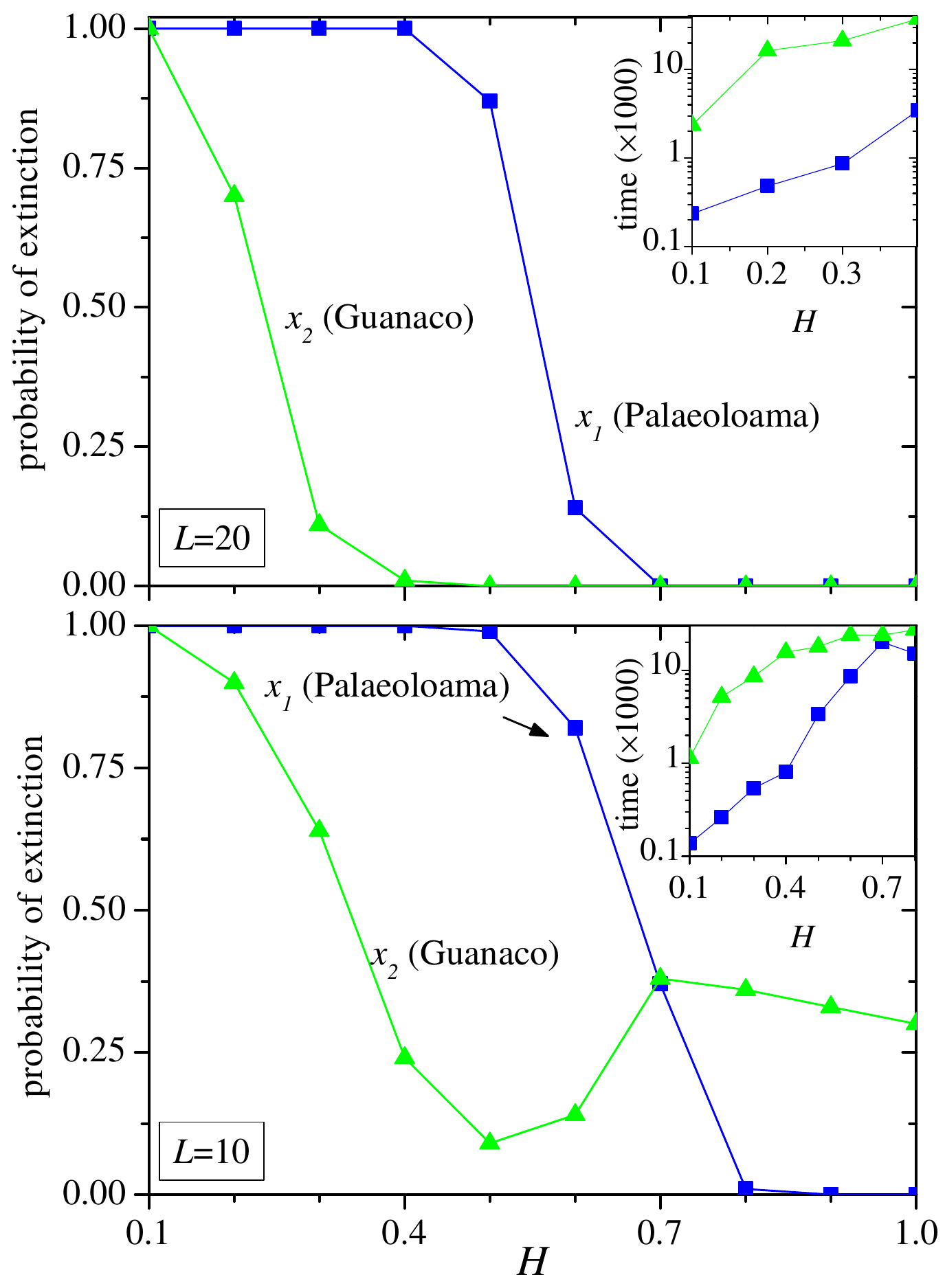}
\caption{Probability of extinction of the two herbivores (palaeolama and guanaco, in absence of humans) as a function of the available habitat $H$, for grids of $20 \times 20$ and $10\times 10$ patches. The quantities shown are averages over $100$ runs, for each value of $H$. Each run was allowed to proceed for a maximum of $60,000$ time steps. Other parameters as in Table 1. Insets: Mean time to extinction as a function of $H$ (semi-log plots).}
\label{fextSH}
\end{figure}

\begin{figure}[t]
\centering 
\includegraphics[width=\columnwidth, clip=true]{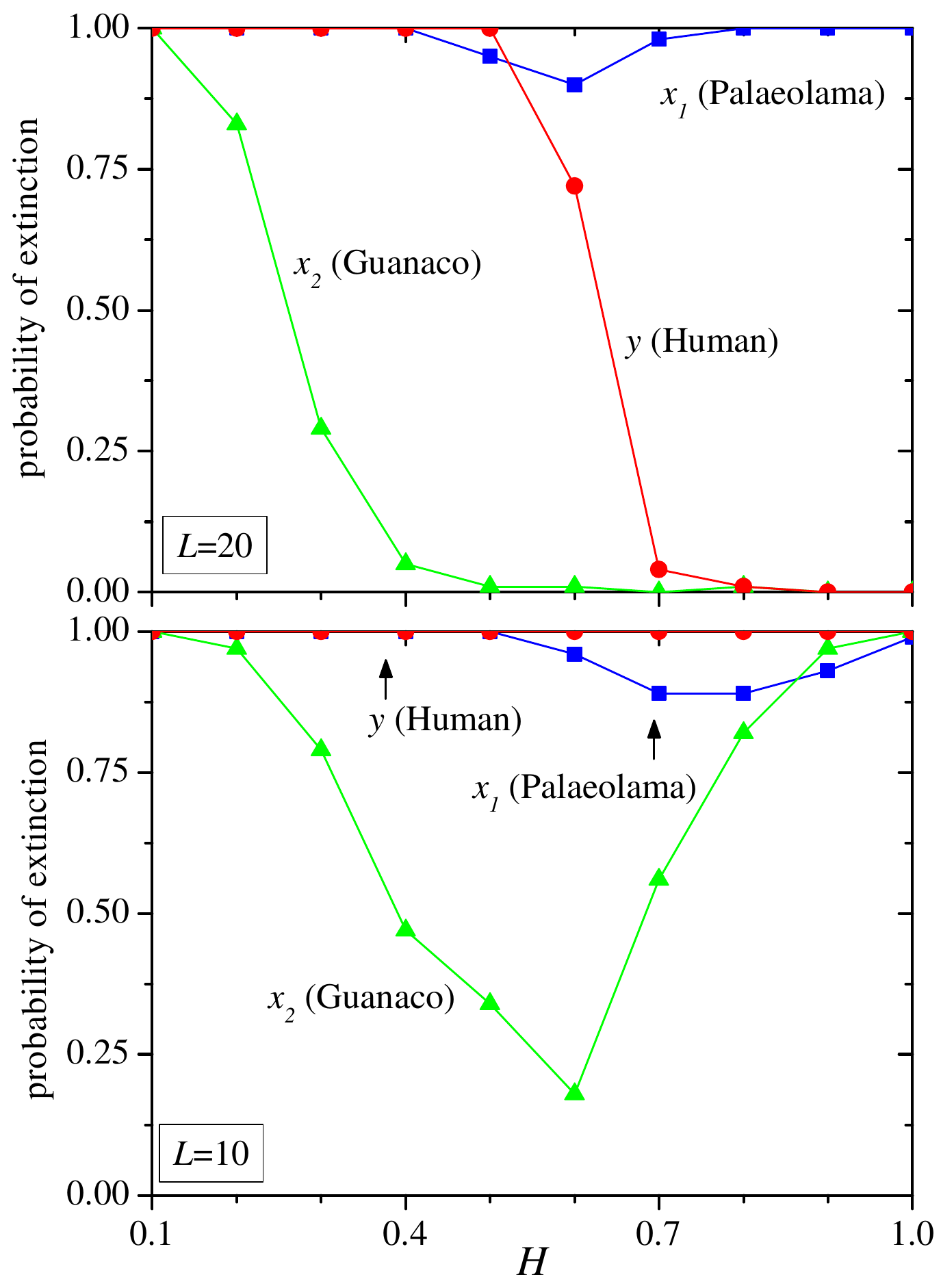}
\caption{Probability of extinction of the two herbivores (palaeolama and guanaco, with human invaders) as a function of the available habitat $H$, for grids of $20 \times 20$ and $10\times 10$ patches. The quantities shown are averages over $100$ runs, for each value of $H$. Each run was allowed to proceed for a maximum of $60,000$ time steps. Other parameters as in Table 1.}
\label{fextCH}
\end{figure}

\subsection{The role of fluctuations in small systems}

As mentioned above, fluctuations can be at least partially responsible for extinctions. While this matter is generally ignored in the study of physical systems that evolve following rules similar to the ones analyzed here, it may be a very relevant fact of the dynamics of ecosystems. The reason for this is that the role of fluctuations is mainly controlled by the size, or any other extensive property of the system. Small populations are clearly subject to the risk of large fluctuations and their consequences. A full mathematical description of the role of fluctuations in the system under study could be done in terms of a Master Equation formulation of the stochastic model and its behavior for ``mesoscopic'' systems (see e.g.~\citet{risau2007},~\citet{abramson2008}). This would take us away from the point of the present study, and we prefer to save it for a future analysis. In any case, the size dependence analysis can be carried out numerically with the tools already described, and in the present section we show such results.

Let us still consider only the strong hierarchy competition scenario, the one that we are calling guanaco-palaeolama. The phenomenon of extinctions due to fluctuations in the absence of hunter-gatherers, presented in the top panel of Fig.~\ref{L10}, is analyzed from a statistical point of view in Fig.~\ref{fextSH}. We show the probability of extinction measured on a set of 100 repetitions of the dynamics, as a function of the available fraction of habitat $H$. 

The system shown in Fig.~\ref{fextSH} (top) is large enough as to display the typical behavior expected for infinite systems (see for example the review book by~\citet{tilman97}). When $H$ is large, the available space is enough for the coexistence of the two competitors, with no extinctions. A reduction of the available habitat produces, firstly, a sharp increase of the probability of extinction of the superior competitor, as expected. Further reduction of $H$ drives also the guanacos to more and more frequent extinctions. Note that coexistence is impossible for $H\lesssim 0.4$, which coincides with the percolation transition of the destroyed patches~\citep{bunde91}. In other words, there is a wide range of $H$, corresponding to a habitat partially destroyed and fragmented, where only the inferior herbivore persists.

The system shown in the bottom panel of Fig.~\ref{fextSH} is smaller than the one just described, and an additional regime can be seen. Observe that there is a reduced probability of coexistence in the (realively pristine) habitats corresponding to $H\gtrsim 0.8$, compared to the larger system. The reason for this is a combination of the lower mean value of the population of guanacos with respect to palaeolamas (observe the time series shown in Fig.~\ref{L10}, corresponding to $H=0.9$) and the large fluctuations due to the very small system size. Now, what happens when less space is available in this scenario? As expected, the superior competitor experiences an increase in its probability of extinction (as in the larger system just described). The disappearance of the palaeolama allows the guanacos to thrive, and we see less and less extinctions of the lower competitor in a scenario of competitive exclusion at intermediate values of $H$. Finally, of course, when $H$ becomes sufficiently small ($H\approx 0.4$), the unavailability of patches start to make an impact on the guanaco population.

We have also observed that, in the situations where both species suffer extinction, the palaeolamas disappear much faster (see the semilog scale) than guanacos, as shown in the insets of Fig.~\ref{fextSH}. This, again, reflects the fact that the superior competitor is the most affected by the destruction of habitat.

Let us turn now to the corresponding phenomena when a small population of human hunter-gatherers is introduced in the system. Figure~\ref{fextCH} shows the system with the three species. The initial condition contains 5\% of the patches occupied by humans, representing an invading population. Even such a small population modifies strongly the survival probability of the herbivores just discussed, especially for the larger values of $H$ (the less destroyed habitats). Observe that, in both system sizes, the palaeolama is particularly affected by the presence of the hunter-gatherers. While in the absence of humans palaeolama was able to thrive with a very small probability of extinction ($H\approx 0.8$-$1.0$, Fig.~\ref{fextSH}), we see here the opposite situation: the probability of extinction is close to 1. Guanacos also experience a strong increase in their probability of extinction, most notable in the smaller system. It is remarkable that this dire fate of the superior competitor happens even when the pressure from hunting---and/or scavenging---is less than on the inferior (see Table 1: $\mu_{x_1}<\mu_{x_2}$). In other words, this is not an extinction due to an overkill of the superior competitor, as in one of the favored hypotheses of the extinction of megafauna \citep{martin1967,alroy2001}. They are simply more vulnerable to both hunting and habitat destruction (as seen in Fig.~\ref{fextSH}). 

\begin{figure}[t]
\centering 
\includegraphics[width=\columnwidth, clip=true]{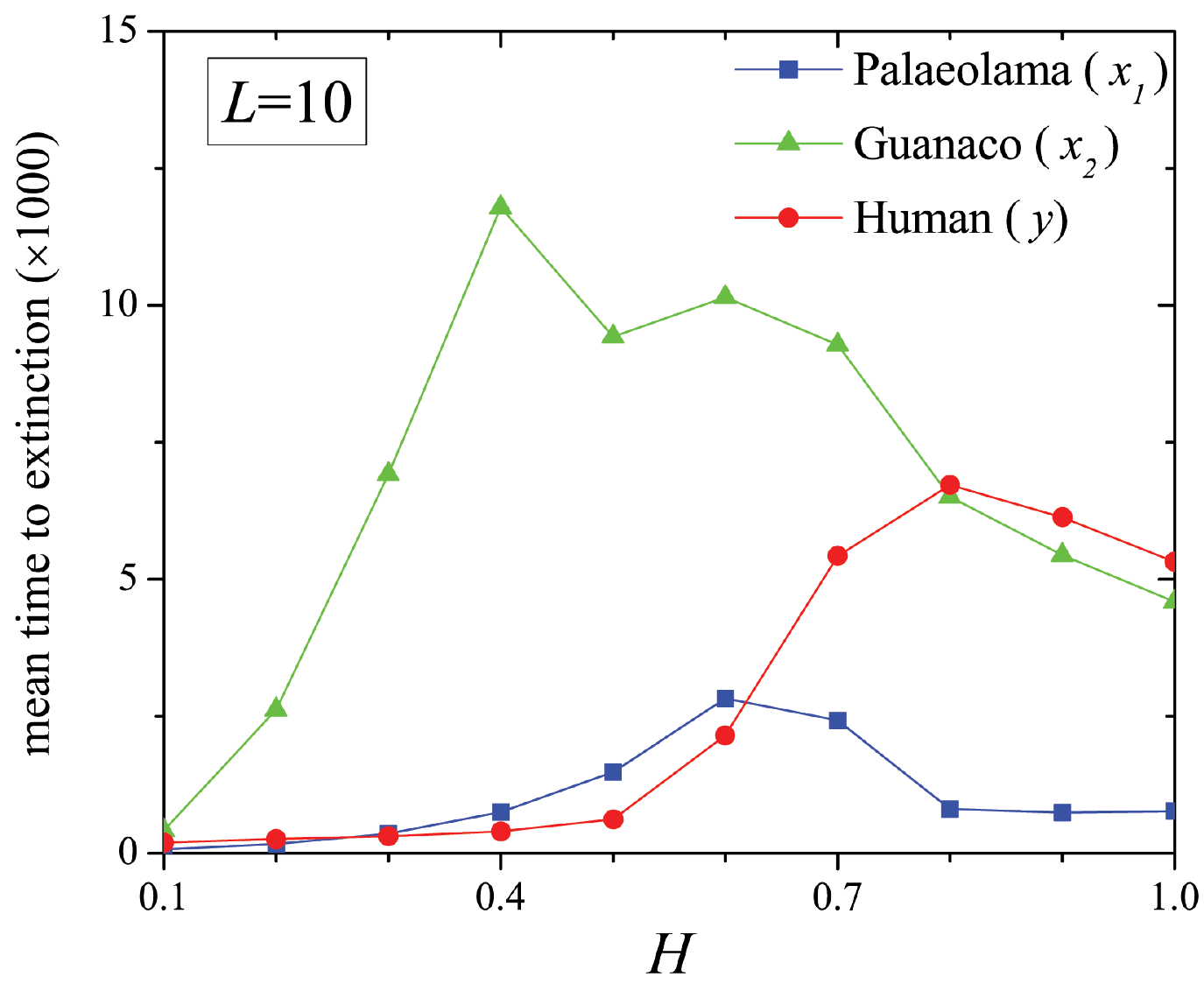}
\caption{Mean time to extinction of the three species as a function of the available habitat $H$. The quantities shown are averages over $100$ runs, for each value of $H$. Each run was allowed to proceed for a maximum of $60,000$ time steps. Other parameters as in Table 1.}
\label{text}
\end{figure}

Strikingly, these highly probable extinction events happen even when the probability of extinction is very high also for the humans. The reason for this is the sequence of these events. Palaeolama is driven to extinction very fast, and the hunter-gatherers persist until they exhaust the availability of the other prey, the inferior competitor. We can see this in Fig.~\ref{text}, which plots the average time to extinction of the three species corresponding to the scenarios of Fig.~\ref{fextCH} (bottom). We see that, as in the case of destruction of habitat, the presence of a predator also affects the superior competitor in a stronger way.

On the other hand, Fig.~\ref{fextCH} shows that for smaller values of $H$ the behavior is qualitatively similar to the situation without humans. The driving force of the fate of the system when $H$ is smaller than around $0.5$ is again the destruction of habitat, which overwhelms the presence of predators. And, again, the superior competitor is the one with a higher probability of extinction.

\section{Discussion}

\begin{figure}[t]
\centering 
\includegraphics[width=\columnwidth, clip=true]{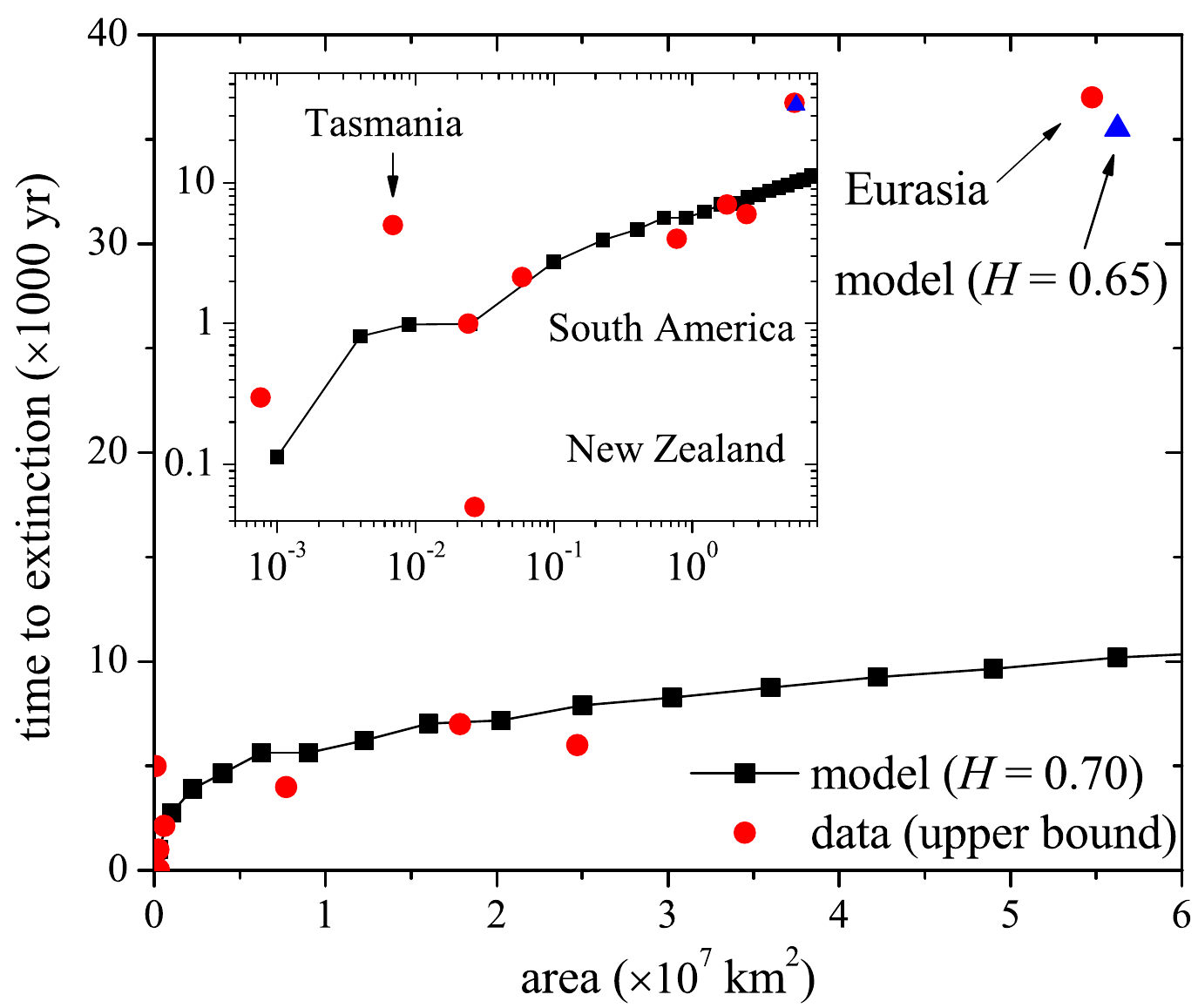}
\caption{Time to extinction of the megafauna after the arrival of human hunters, as a function of the area of the system. The model results correspond to averages of 1000 realizations run on square systems with randomly distributed destruction (with values of $H$ as indicated). The real world data correspond to the most recent date for each land mass as compiled by \citet{araujo2015}. From smallest to largest these are: Wrangel Island, Tasmania, the Caribbean, New Zealand, Madagascar, Australia, South America, North America, Eurasia. Other parameters as in Table 1. Inset: same data in a log-log plot.}
\label{extincion}
\end{figure}

We have analyzed a minimal model of a system showing the extinction of megafauna under the influence of three main factors. On the one hand, the system is composed of two species in competitive coexistence, hierarchically sharing a common resource. On the other hand, the system also includes a predator, representing human hunter-gatherers in the first stages of an invasion of the ecosystem. Finally, the model considers the deterioration of the habitat, representing global environmental changes. These two last factors are thought to have played a role in the extinction of megafauna at the end of the Pleistocene, in South America and in practically every other region of the world. We have characterized the actors (see Figs.~\ref{milodon} to~\ref{L20}) as humans, guanacos, palaeolama and mylodon, but certainly the same conclusions apply to any trophic web of similar characteristics. It is important to emphasize that the humans have been modeled as strictly predators, occupying regions containing game. The transition to sedentary human lifestyle has not been considered.

Our contribution has the purpose of providing a mathematical framework in which different possible ecological scenarios can be studied and their consequences contrasted with the paleontological evidence. This first step already shows several interesting results that are worth discussing.

There are a number of relevant questions about the system that can be addressed in the light of the kind of model we have analyzed, as follows (see~\cite{monjeau2015}, this issue):
\begin{enumerate}
\item If the habitat would have been optimal when humans arrived, would the extinction have happened anyway? 
\item If humans had not arrived, would the extinction have happened? 
\item How does the area of the landscape affect the outcome? 
\item How do other details of the natural history of the fauna (body size, reproduction rate, competitive ability, etc.) affect the outcome? 
\end{enumerate}
Within the framework of this first step, let us discuss briefly our contribution in the direction of answering this kind of questions.

Regarding the first matter (a), about the effect of hunting in a pristine habitat, Fig.~\ref{fextCH} shows that extinction of megafauna happens even in the optimal habitat ($H=1$) for any system area. The large one in particular ($L=20$, equivalent to an infinite area in the present context),  exhibits the extinction of megafauna together with the coexistence of the hunters with the extant herbivore. 

It is also interesting to note that the extinction of the megafauna occurs for the whole range of habitat availability. In a situation of coexistence of the two herbivores, the introduction of humans very quickly drives the superior one (arguably the largest one) to extinction, while a coexistence of the hunter-gatherers with the smaller prey ensues when $H>0.5$. We stress once again that this shows a possible scenario of extinction of megafauna without the overkill hypothesis extensively studied in recent years. 

In addition, we see that there is a regime of extinction of humans while game is still available, as can be seen in Fig.~\ref{fextCH} for intermediate values of $H$. This reflects the fact that the inferior herbivore is a better colonizer, and is best fit to survive in a deteriorating habitat.

A comparison of Fig.~\ref{fextCH} with Fig.~\ref{fextSH}, in turn, provides some insight into the second question (b): extinctions due to just environmental causes. Fig.~\ref{fextSH} shows that, as expected, the megafauna can indeed be driven to extinction by habitat deterioration. The phenomenon affects steeply the probability of extinction of the superior competitor when the destroyed habitat approaches the percolation transition in the system (about $H=0.4$). Bear in mind that the destruction of habitat implemented here is uniform and random; some details of this transition might depend on the particular spatial arrangement of the landscape, but the general conclusion would still hold.

Habitat destruction is known to be detrimental for the higher competitor in a hierarchy. This can be understood in the sense that the lower competitor, in a situation of coexistence, needs some advantage to compensate for the asymmetric interaction: being a better colonizer, for example. This leaves the superior one (typically the largest one) in harms way when the habitat reduces or fragments. We have shown that the results may differ from this due to fluctuations, a phenomenon related to question (c). 

The effect of the area can be preliminary addressed by comparing top and bottom panels of Figs.~\ref{fextSH} and~\ref{fextCH}. Here we see that the main effect of the system size reduction is on the survival probability of the inferior competitor. It is an effect driven by fluctuations, which are particularly strong in smaller systems. Such systems can see the extinction of the inferior competitor when the habitat is almost pristine (Fig.~\ref{fextSH}). A reduction of the available habitat, nevertheless, is more strongly felt by the superior competitor. With a reduced competitive pressure, the inferior competitor correspondingly reduces its probability of extinction. In the presence of humans the situation changes considerably, especially for the more pristine habitats. And, as a result, it is again the superior competitor the one that suffers most. They are driven to extinction with a high probability for small and large systems alike. 

We have also seen that the time to extinction is a widely distributed magnitude. In particular, longer extinction times (both in the presence and without humans) correspond to larger systems. This fact has a well-known correlate in the real world, as extinctions in continents took typically longer than in large islands, which in turn took longer than in small islands \citep{araujo2015}. On the light of this, it is interesting to inquiry the effect of the area on a global scale, comparing the history of several landmasses. We make an assessment of the strength of our model to deal with this matter in Fig.~\ref{extincion}, where we plot the time to extinction of the megafauna after the injection of human hunter-gatherers. The red circles correspond to the most recent time to extinction for each land mass, as compiled by \citet{araujo2015}. The black squares (and the isolated blue triangle close to Eurasia) are results from our model, corresponding to ensemble averages (1000 realizations) run on square systems from $2\times 2$ to $200\times 200$. The values of $H=0.7$ (and $H=0.65$ for Eurasia) are the ones that best adjust the data. The temporal scale in the model was taken to be 1 year corresponding to 1 simulation step. This is a reasonable choice for a metapopulation of large vertebrates. It is motivated by their demographics and natural history, which is strongly influenced by seasonality, and accounts, for example, for the eventual spread to neighboring patches or not surviving winterkill. The size scale was calibrated to fit the data, resulting in patches of 2500 km$^2$, which is also very reasonable. 

The results of the model show a reasonable agreement with the data. There are several exceptions worth of note. Firstly, we see that the situation in Eurasia is strikingly different: the point doesn't follow the smooth tendency of the rest of the landmasses. It has been argued that this reflects the fact that the giant continent was occupied more slowly by humans, with difficulties arising from the extension and the need to construct new niches for survival \citep{monjeau2015}. These, and probably many other facts, contributed to a longer survival of the megafauna in Eurasia. In the model, the situation is equivalent to a smaller value of $H$, and we have found that $H=0.65$ is a close match. Given that $H$ condenses several aspects of the environment, its interpretation should not be taken strictly as the actual availability of habitat. Further refinements of the model will attempt to better characterize these matters. Secondly, on the smallest size scales we see strong fluctuations of the time to extinction as a function of area (e.g. the cases of Tasmania and New Zealand). This is also to be expected since, as we argued before, smaller systems are prone to stronger stochastic fluctuations. Individual realizations (as the ones actually observed in the real world) are expected to reflect this. Aside from these details, the model reproduces the trend of the data across more than three orders of magnitude of the area.

The last question (d), regarding other details of the natural history of the fauna, has been assessed very indirectly in this simple model, because the many peculiarities of each species are drastically condensed into a few parameters controlling colonization, extinction and predation rates. Nevertheless, we have considered two types of competitive interactions (the strong and the weak ones, related to herbivores with different overlap of diets in section 3.1). We have seen that the same qualitative results are obtained in both cases: the extinction of the superior competitor shortly after the introduction of humans. 

It is remarkable that our simple model is able to capture both qualitatively and quantitatively several apects of such a complex ecosystem. The results found in this work are encouraging to attempt the modelization of more realistic scenarios, yet within the conceptual simplification of model~(\ref{transmodel1})-(\ref{transmodel3}). This will be done in two main directions. On the one hand, the spatial landscape can be easily modified to fit the shape and habitats of a continent. This, in turn, will allow for a more realistic implementation of the human invasion and the progress of its impact on the native fauna. On the other hand, the biotic variables (especially the trophic structure of the system) can be made more complex by resorting to known facts about the Pleistocene fauna. In all cases, let us mention again that the role of fluctuations and spatial correlations needs to be thoroughly assessed. 

As concluded in \citet{monjeau2015}, the extinction of megafauna due to anthropic or climatic driven effects cannot be absolutely proven in all scenarios due to the lack of field evidences and to the synergy of multiple causes. As \citet{johnson2002} said, paleontological and archaeological data may not be able to provide conclusive arguments over the causes of megafauna extinctions, at least in the near future. The present work attempts to provide indicators and a solid mathematical formalism to motorize the discussion about the plausibility of the hypotheses.


\appendix
\section{Mean field approximation}
\setcounter{figure}{0}  

An approximate analytical version of the stochastic model formulated in Section 2 can be cast in the form of a set of differential equations as follows (see \cite{laguna2015} for further details). Let $x_1$, $x_2$ and $y$ be the fractions of occupied patches, as before, but let them be differentiable functions of time. If we ignore any explicit spatial dependence of the variables, a mean field model is given by the equations:
\begin{align}
\frac{dx_1}{dt} &= c_{x_1} x_1 (H-x_1) - e_{x_1} x_1 -\mu_{x_1} x_1 y, \label{mfx1} \\
\frac{dx_2}{dt} &= c_{x_2} x_2 (H-x_1-x_2) - e_{x_2} x_2 - c_{x_1} x_1 x_2 - \mu_{x_2} x_2 y, \label{mfx2} \\
\frac{dy}{dt} &= c_y y (x_1+x_2-x_1 x_2) - e_y y. \label{mfy}
\end{align} 
Being spatially implicit, such a model ignores the short range correlations between occupied patches that arise from the local and first-neighbor dynamics of the metapopulations. Because of this, and also because of the role of fluctuations, the exact values of parameters giving a specific result in the stochastic model do not need to give the same in the deterministic one. Nevertheless, some global features are still captured by such a mean field model, and the analysis is somewhat simplified thanks to the possibility of an analytical treatment. In particular, the robustness of the results presented in Section 3 with respect to the choice of parameters can be assessed. Before showing these results, let us briefly discuss the terms appearing in Eqs.~(\ref{mfx1}-\ref{mfy}).

As in the stochastic model, occupation grows at a rate proportional to the fraction of available patches and to the already occupied ones. The fraction of the patches available for colonization for each species is given by the terms in parentheses in Eqs.~(\ref{mfx1}-\ref{mfy}). For the herbivores these terms reflect the reduction of available space through the parameter $H=1-D$ and the already occupied patches of each species. They also contain the asymmetrical nature of the competition in the fact that $x_2$, the inferior competitor, has a further reduction corresponding to the space occupied by the superior species $x_1$. The hunters, in turn, can only colonize patches that already contain game: these are the sum $x_1\!+\!x_2$ minus the ones that, on average, are occupied by both, $x_1x_2$, which are counted twice in the sum. The extinction and predation terms complete the equations, and their interpretation is straightforward. Observe only an additional extinction term in Eq.~(\ref{mfx2}) corresponding to the competitive displacement by the superior herbivore (in the case of strong interaction, as discussed in Section 2). This happens at the rate of colonization of $x_1$, and it is proportional to the product of both species.

\begin{figure}[t]
\centering 
\includegraphics[width=\columnwidth, clip=true]{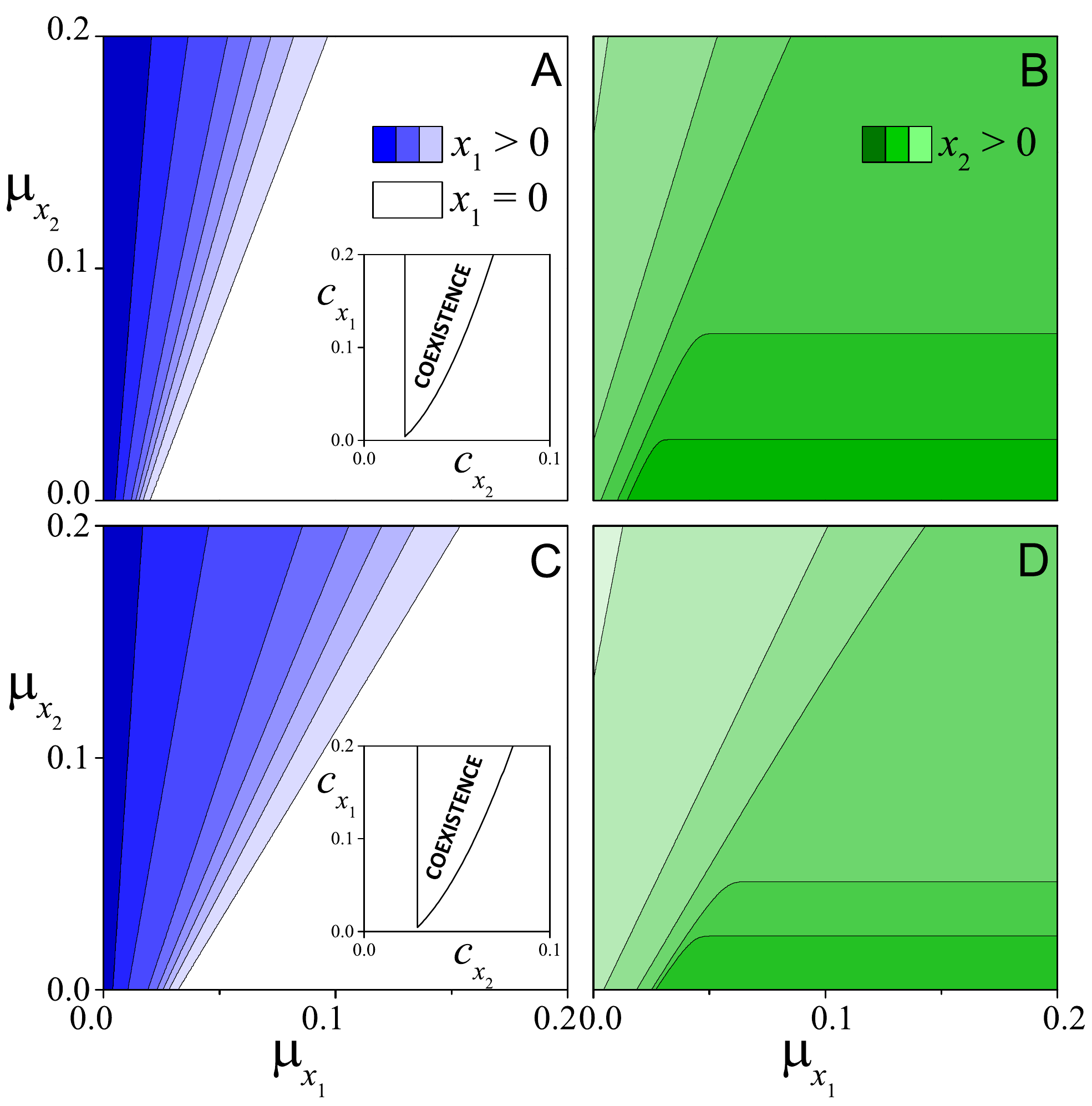}
\caption{Extinction and persistence of the herbivores under a variety of scenarios in the presence of hunters. Left panels (blue on-line) correspond to $x_1$ (palaeolama); right panels (green on-line) to $x_2$ (guanaco). The darker shades represent greater occupied fractions of the system (the exact values are not important); white is extinction. The insets show the regions of coexistence of $x_1$ and $x_2$ in the absence of $y$. Parameters are: panels A and B,  $H=0.9$, $c_{x_1}=0.03$, $c_{x_2}=0.05$, $c_y=0.03$, $e_{x_1}=0.02$, $e_{x_2}=0.008$, $e_y=0.01$; panels C and D, $H=0.7$, $c_{x_1}=0.04$, $c_{x_2}=0.05$, $c_y=0.04$, $e_{x_1}=0.02$, $e_{x_2}=0.008$, $e_y=0.01$.}
\label{meanfield}
\end{figure}

The system (\ref{mfx1}-\ref{mfy}) allows the exploration of the range, in the multidimensional space of all the parameters, where the extinction of the megafauna (species $x_1$) follows after the introduction of hunters. We have found that this region is broad, giving a reassurance about the robustness of the extinction scenarios exemplified in Section 3. We have chosen a limited analysis to show here, summarized in Fig.~\ref{meanfield}. First, we need a set of parameters that gives coexistence of the herbivores in the absence of predators. The conditions of such coexistence are easy to derive from Eqs.~(\ref{mfx1}-\ref{mfy}), requiring that the equilibriums of $x_1$ and $x_2$ are positive when $y=0$:
\begin{align}
0&< H-\frac{e_{x_1}}{c_{x_1}}, \\
0&<\frac{e_{x_1}}{c_{x_1}}+\frac{e_{x_1}-e_{x_2}}{c_{x_2}} - \frac{c_{x_1}}{c_{x_2}}H.
\end{align}
This conditions, of course, depend on the parameter $H$. In Fig.~\ref{meanfield} the insets show the region of coexistence for two values of $H$, in the space spanned by the colonization rates, keeping the other parameters fixed. The fact that one of the boundaries of the region of coexistence is a straight vertical line reflects the hierarchy: the superior competitor does not feel the presence of the inferior one. This, on the other hand, needs to be a better colonizer in order to persist under the asymmetric competition, and the curved shape of the coexistence region shows this.

Given coexistence in the absence of hunters, the remaining plots of Fig.~\ref{meanfield} show the conditions for extinction in the space of hunting pressure, keeping the other parameters constant. We chose this representation to show that the megafauna is driven to extinction (white regions) even in scenarios where the inferior herbivore is the preferred prey. This happens above the identity diagonal $\mu_{x_2}=\mu_{x_1}$. Certainly, there is also extinction if the superior one is the preferred prey (but not always: see the region of coexistence of the three species located in the region of small $\mu$'s). These are \emph{overkill} scenarios. We want to stress, however, the possibility of extinction of the megafauna in the absence of overkill. The reason for this is the same already argued in Section 3. The persistence of the superior competitor is more fragile regarding both threats: the deterioration of habitat (through $H$, a point already noted by~\citet{tilman94}) and the pressure of hunting. This fragility is due to the fact that \emph{for coexistence} the inferior competitor needs some survival advantage, and this puts the megafauna in harm's way. The parameters chosen in Table 1 correspond to this scenario of coexistence in the pre-human situation, better colonization performance of the smaller herbivore, and an increased hunting pressure on them. Their values reflect a reasonable situation that, together with the robustness of the results with respect to the parameters, provide verisimilitude to our analysis.

\section{Simulation details}

A square grid of $L\times L$ sites is set up with a fraction $D$ of them marked as destroyed and unsuitable for occupation. Boundary conditions are impenetrable barriers (effectively implemented by destroying all the patches in the perimeter). Initial populations of herbivores are distributed uniformly at random on the undestroyed space. After a transient a population of hunters is distributed at random on suitable sites. Let $x_i(j,t)\!\in\!\{0,1\}$ and $y(j,t)\!\in\!\{0,1\}$ be the occupation of site $j$ by the corresponding species at time $t$. (With $i=1,2$ for the superior and inferior herbivores respectively, and using the single spatial label $j$ for brevity, even though the actual system is two-dimensional.) The variables used in the Results are related to these as: $x_i(t) = \sum_j x_i(j,t)/L^2$ and $y(t) = \sum_j y(j,t)/L^2$, summing over the whole system.

At each time $t$ all non-destroyed patches are visited. Based on their occupation state the dynamical rules are applied as follows.

\textbf{Colonization} 
\begin{itemize}
\item If $x_1(j,t)\!=\!0$, set $x_1(j,t\!+\!1)\!=\!1$ with probability $1\!-\!(1\!-c_{x_1})^{v_1(j,t)}$, where $v_1(j,t)$ is the number of nearest neighbors of $j$ already occupied by species 1 at time $t$. This takes into account the chances of colonization arriving from any of the neighbor patches.
\item If $x_2(j,t)\!=x_1(j,t)\!=\!0$, set $x_2(j,t\!+\!1)\!=\!1$ with probability $1\!-\!(1\!-c_{x_2})^{v_2(j,t)}$, where $v_2(j,t)$ is analogous to $v_1(j,t)$, for species 2.
\item If $y(j,t)=0$ and: $x_1(j,t)\!=1$ or $x_2(j,t)\!=\!1$, set $y(j,t\!+\!1)\!=\!1$ with probability $1\!-\!(1\!-c_y)^{v_y(j,t)}$, where $v_y(j,t)$  is analogous to $v_1(j,t)$, for the hunters.
\end{itemize}

\textbf{Extinction and predation}
\begin{itemize}
\item If $x_1(j,t)\!=\!1$, set $x_1(j,t+1)\!=\!0$ with probability $e_{x_1}+\mu_{x_1}\,y(j,t) $. This combines the effects of extinction and predation when the hunters are present (the same is done for species 2, in the following item).
\item If $x_2(j,t)\!=\!1$, set $x_2(j,t+1)\!=\!0$ with probability $e_{x_2}+\mu_{x_2}\,y(j,t) $. 
\item If $y(j,t)\!=\!1$, set $y(j,t\!+\!1)\!=\!0$ with probability $e_y$.
\end{itemize}

\textbf{Competitive displacement}
\begin{itemize}
\item If $x_1(j,t)=x_2(j,t)=1$, set $x_2(j,t+1)=0$ with probability $c_{x_1}$. This is implemented in the strong competition case only.
\end{itemize}

After visiting all the sites of the system, update the state of the grid and iterate. Since the update is in parallel for all the sites of the grid, the order of application of the dynamical rules is not important. The results presented in Figures 2, 3, 4 and 5 are realizations of the algorithm just described. Figures 6, 7, 8 and 9 consist of ensembles of equivalent realizations of the stochastic model, run up to a prescribed maximum time. At the end of this one or more species may have become extinct. Averages of the reported quantities (probability of extinction and mean time to extinction) are performed over this ensemble.

\section*{Acknowledgements}
We acknowledge financial support from several sources: CONICET (PIP 112-201101-00310), Universidad Nacional de Cuyo (06/C410), ANPCyT (PICT-2011-0790 and PICT-2014-1558), UNRN 40/B134, BC IPM 2013. AM was funded by CONICET and CNPq for a research stay at the Universidade Federal do Rio de Janeiro (PVE 400363/2014-3) and acknowledges colleagues at the Fundaci\'on Bariloche. We thank Bernardo Araujo for the preparation of Figure 1.

\bibliographystyle{elsarticle-harv} 
\bibliography{megafauna}






\end{document}